\documentclass[twocolumn,floatfix,superscriptaddress]{revtex4}
\usepackage{graphicx}
\usepackage{amsmath}
\usepackage{amssymb}
\usepackage{bm}

\usepackage{color}

\begin{document}

\title{Emergence of massless Dirac fermions in graphene's Hofstadter butterfly at switches of the quantum Hall phase connectivity}
\author{M. Diez}
\affiliation{Instituut-Lorentz, Universiteit Leiden, P.O. Box 9506, 2300 RA Leiden, The Netherlands}
\author{J. P. Dahlhaus}
\affiliation{Department of Physics, University of California, Berkeley, California 95720, USA}
\author{M. Wimmer}
\affiliation{Kavli Institute of Nanoscience, Delft University of Technology, P.O. Box 5046, 2600 GA Delft, The Netherlands}
\author{C. W. J. Beenakker}
\affiliation{Instituut-Lorentz, Universiteit Leiden, P.O. Box 9506, 2300 RA Leiden, The Netherlands}

\date{January 2014}
\begin{abstract}
The fractal spectrum of magnetic minibands (Hofstadter butterfly), induced by the moir\'{e} superlattice of graphene on an hexagonal crystal substrate, is known to exhibit gapped Dirac cones. We show that the gap can be closed by slightly misaligning the substrate, producing a hierarchy of conical singularities (Dirac points) in the band structure at rational values $\Phi=(p/q)(h/e)$ of the magnetic flux per supercell. Each Dirac point signals a switch of the topological quantum number in the connected component of the quantum Hall phase diagram. Model calculations reveal the scale invariant conductivity $\sigma=2qe^2/\pi h$ and Klein tunneling associated with massless Dirac fermions at these connectivity switches.
\end{abstract}
\maketitle

The quantum Hall effect in a two-dimensional periodic potential has a phase diagram with a fractal structure called the ``Hofstadter butterfly''  \cite{Hof76,Osa01}. In a 2013 breakthrough, three groups reported \cite{Pon13,Dea13,Hun13} the observation of this elusive structure in a graphene superlattice, produced by the moir\'{e} effect when graphene is deposited on a boron nitride substrate with an almost commensurate hexagonal lattice structure. It was found that the magnetic minibands repeat in a self-similar way at rational values $\Phi/\Phi_0=p/q$ of the flux $\Phi$ through the superlattice unit cell, with $p,q$ integers and $\Phi_0=h/e$ the flux quantum.

A central theme of studies of the Hofstadter butterfly is the search for flux-induced massless Dirac fermions \cite{Ram85,Hou09,Ger10,Del10,Rhi12}. It turns out that in the graphene superlattice only the zero-field Dirac cones are approximately gapless \cite{Sac11,Ort12,Yan12,Kin12}, while the flux-induced Dirac cones are gapped \cite{Che13}. Generically, Dirac fermions in the Hofstadter butterfly are massive.

\begin{figure}[tb]
\centerline{\includegraphics[width=0.9\linewidth]{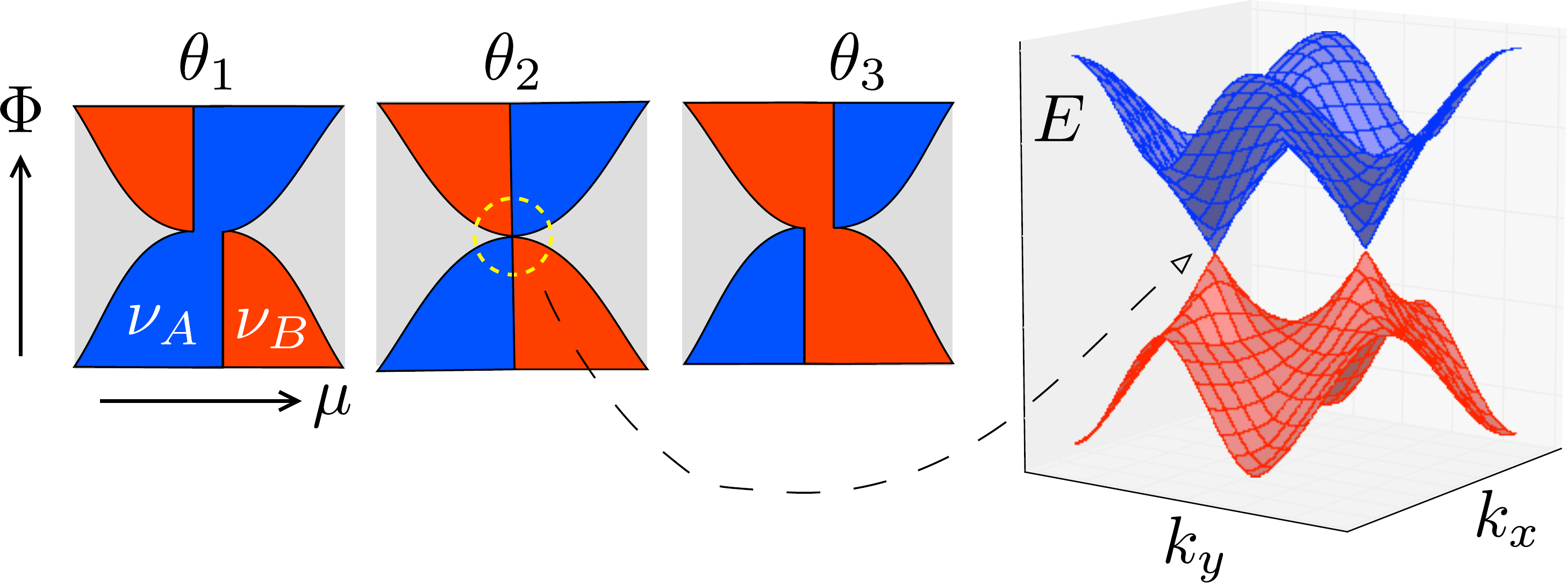}}
\caption{Schematic illustration of a connectivity switch in the quantum Hall phase diagram. Upon variation of a control parameter $\theta$ the connected component switches from topological quantum number $\nu_A$ to $\nu_B$. At the transition a singular point appears in the phase boundary (encircled), associated with gapless Dirac cones in the Brillouin zone (right-most panel). 
}
\label{fig_transition}
\end{figure}

Here we show that massless Dirac fermions do appear at singular points in the quantum Hall phase diagram, associated with a switch of the phase connectivity upon variation of some control parameter. (See Fig.\ \ref{fig_transition}.) Any experimentally accessible quantity that couples to the superlattice potential can play the role of control parameter, in what follows we will consider the angle $\theta$ of crystallographic alignment between graphene and substrate. We find that the phase boundaries separating regions of distinct Hall conductance $\sigma_{xy}=\nu e^{2}/h$ rearrange their connectivity upon variation of $\theta$, switching the connected component of the phase diagram from $\nu$ to $\nu\pm 2q$. In the magnetic Brillouin zone this transition produces a pair of $q$-fold degenerate conical singularities (Dirac points), with massless Dirac fermions as low-energy excitations.

We base our analysis on the moir\'{e} superlattice Hamiltonian of Wallbank {\em et al.} \cite{Wal13}. Starting point is the Dirac Hamiltonian of graphene \cite{Cas09,Kat12},
\begin{equation}
H_{0}=v[\bm{p}-e\bm{A}(\bm{r})]\cdot\bm{\sigma}+V(\bm{r}),\label{H0def}
\end{equation}
for conduction electrons near each of two opposite corners (valleys) of the hexagonal Brillouin zone \cite{note1}. The Fermi velocity is $v=10^{6}\,{\rm m/s}$ and the lattice constant of the hexagonal lattice of carbon atoms is $a=2.46\,\mbox{\AA}$. The momentum $\bm{p}=-i\hbar\nabla$ in the $\bm{r}=(x,y)$ plane is coupled to pseudospin Pauli matrices $\sigma_{x}$ and $\sigma_{y}$ acting on the sublattice degree of freedom. The real spin plays no role and is ignored \cite{note2}, only the orbital effect of a  perpendicular magnetic field $\bm{B}=B\hat{z}$ is included (via the vector potential $\bm{A}$). The electrostatic potential $V$ is adjustable via a gate voltage. For simplicity we assume that the mean free path for impurity scattering is sufficiently large that disorder effects can be neglected. 

The moir\'{e} effect from a substrate of hexagonal boron nitride (hBN, lattice constant $(1+\delta)a$, $\delta=0.018$, misaligned by $\theta\ll 1$) adds superlattice terms to the Dirac Hamiltonian. The terms that break inversion symmetry are small and we neglect them, following Ref.\ \cite{Abe13}. Three terms remain \cite{Wal13},
\begin{align}
H={}&H_{0} + \hbar vbU_1 f_{+}(\bm{r})+i\xi \hbar vb U_2\sigma_{z} f_{-}(\bm{r})\nonumber\\
& +i\xi \hbar v U_{3}\left(\sigma_{y}\partial f_-/\partial x-\sigma_{x}\partial f_-/\partial y\right),\label{eq:model}
\end{align}
where $\xi=\pm 1$ in the two valleys and
\begin{align}
&f_{\pm}(\bm{r})=\sum_{m=0}^5  (\pm 1)^{m} e^{i\bm{b}_m \bm{r}}=\pm f_{\pm}(-\bm{r}),\label{f1f2def}\\
&\bm{b}_m=\frac{4\pi}{\sqrt{3}a}\hat{R}_{\pi m/3}\left[ 1-(1+\delta)^{-1} \hat{R}_\theta \right] \begin{pmatrix}
0\\
1
\end{pmatrix}.\label{bmdef}
\end{align}
The reciprocal lattice vectors $\bm{b}_{m}$ have length $b\equiv|\bm{b}_{0}|\approx (4\pi/\sqrt{3}a)\sqrt{\delta^2+\theta^2}$ and are rotated by the matrix
\begin{equation}
\hat{R}_{\theta}=\begin{pmatrix}
\cos\theta&-\sin\theta\\
\sin\theta&\cos\theta
\end{pmatrix}.\label{Ralphadef}
\end{equation}
The periodicity of the superlattice is $\lambda=4\pi/\sqrt{3}b\approx a/\sqrt{\delta^2+\theta^2}$.

The terms $U_1$ and $U_2$ in the Hamiltonian \eqref{eq:model} represent a potential modulation, while the term $U_3$ is a modulation of the hopping amplitudes. The coefficients are related by \cite{Kin12,Wal13}
\begin{equation}
\left\{U_1, U_2, U_3\right\} = \frac{E_{0}}{\hbar vb}\left\{\tfrac{1}{2}, -\tfrac{1}{2}\sqrt{3},-(1+\theta^2/\delta^2)^{-1/2}\right\},\label{eq:u0u1u3}
\end{equation}
where $E_{0}$ is an energy scale that sets the coupling strength of graphene to the hBN substrate. We use the estimate $E_{0}=17\,{\rm meV}$ from Ref.\ \cite{Abe13}, corresponding to a ratio $E_{0}/\hbar vb=0.05\,(1+\theta^2/\delta^2)^{-1/2}$.

\begin{figure}[tb]
\centerline{\includegraphics[width=0.8\linewidth]{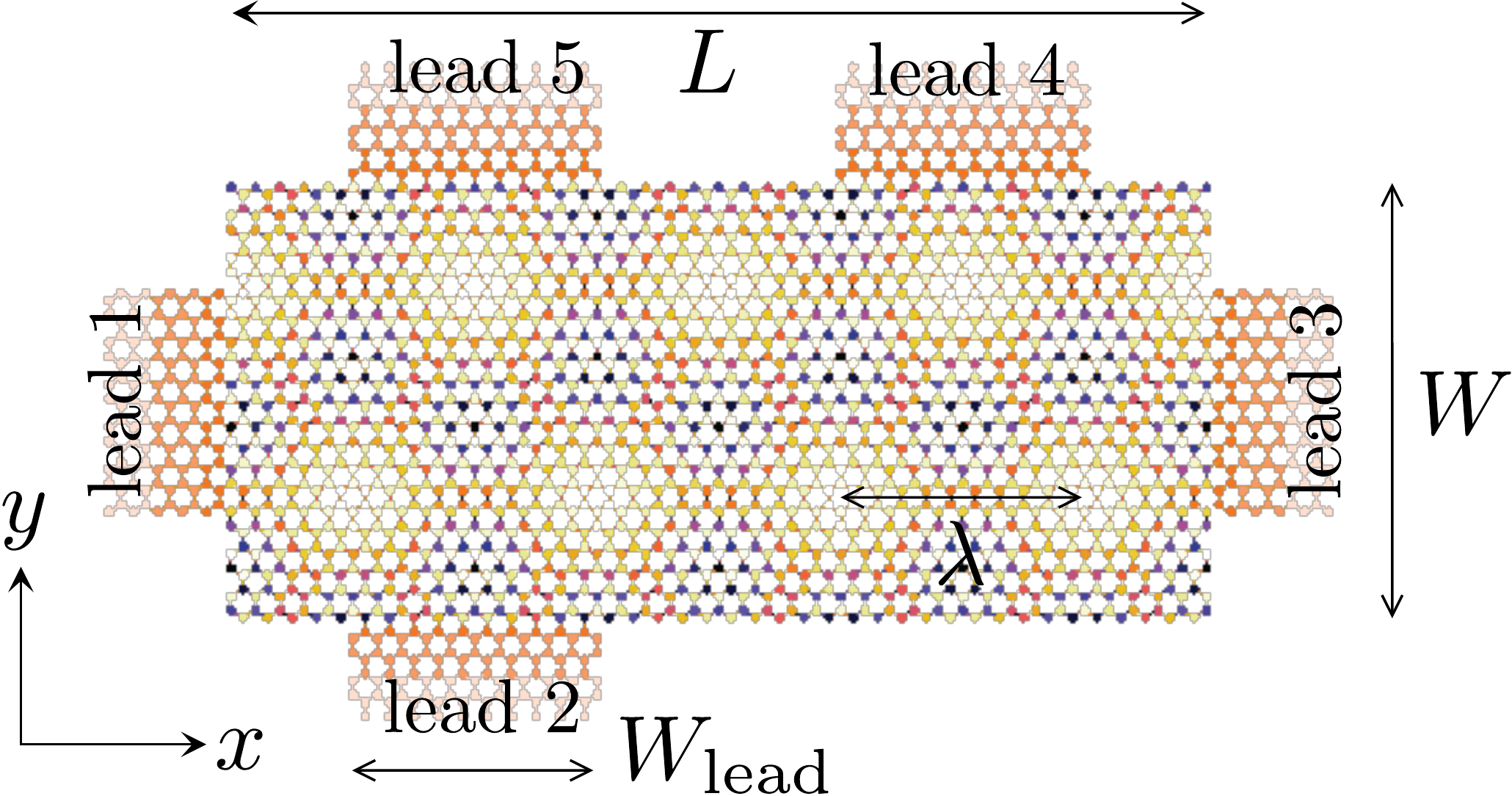}}
\caption{Five-terminal geometry used to calculate the Hall conductivity \eqref{eq:sigmaxy}. The two-dimensional hexagonal lattice of the tight-binding model is shown, with the superlattice potential indicated by colored sites and bonds (not to scale, the actual lattice is much finer). 
}
\label{fig_hallbar}
\end{figure}

\begin{figure*}[tb]
\centerline{\includegraphics[width=1\linewidth]{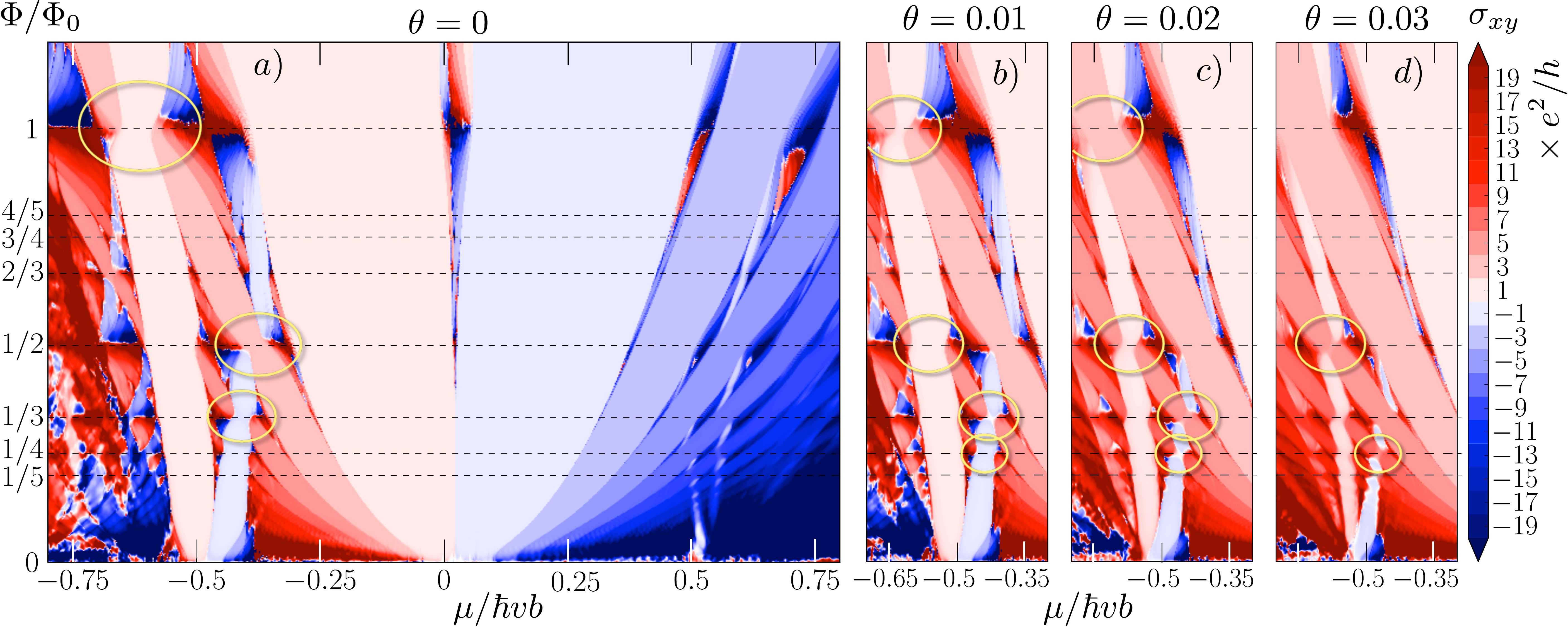}}
\caption{Numerical results for the Hall conductivity of graphene on hBN, calculated in the Hall bar geometry of Fig.\ \ref{fig_hallbar} for the superlattice Hamiltonian \eqref{eq:model}. Panel \textit{a} is for a perfectly aligned substrate, when the flux-induced Dirac cones (encircled) are all gapped. Panels \textit{b,c,d} show the connectivity switches induced by a slight crystallographic misalignment of the substrate (angle $\theta$ in radians). 
}
\label{fig:sigmaxy}
\end{figure*}

We study electrical conduction in the five-terminal Hall bar geometry of Fig.\ \ref{fig_hallbar}, where a current $I$ flows from source 1 to drain 3 while contacts 2, 4, and 5 draw no current. The voltages $V_n$ at these contacts determine the Hall conductivity,
\begin{equation}
\sigma_{xy}=\frac{(V_5-V_2)I}{(V_5-V_2)^2+(W/L)^2(V_5-V_4)^2}.\label{eq:sigmaxy}
\end{equation}
In linear response and at zero temperature the voltage differences are obtained from the scattering matrix $S(E)$ at the Fermi level $E_{\rm F}=0$, which we calculate by discretizing the Hamiltonian \eqref{eq:model} on a tight-binding lattice (hexagonal symmetry, lattice constant $a_{\rm TB}=\lambda/20$). The metallic contacts are modeled by heavily doped graphene leads (infinite length, width $W_{\rm lead}=5\lambda$, potential $V_{\rm lead}=2\,\hbar vb$), without the superlattice ($E_0=0$ in the leads) and without magnetic field. In the superlattice region (length $L=20\lambda$, width $W=5\sqrt{3}\lambda$) we set $V=-\mu$. (The sign of $\mu$ is chosen such that the Fermi level lies in the conduction band of graphene for $\mu>0$ and in the valence band for $\mu<0$.) We calculate $\sigma_{xy}$ as a function of $\Phi$ and $\mu$ using the {\sc kwant} tight-binding code \cite{kwant,note3}. Results are shown in Fig.\ \ref{fig:sigmaxy}.

Panel \ref{fig:sigmaxy}\textit{a} shows the known spectral features of the graphene superlattice \cite{Pon13, Dea13, Hun13,Che13}: A parabolic fan of Landau levels emerging from the primary zero-field Dirac cone of graphene; secondary zero-field Dirac cones centered at $\mu=\pm \hbar vb/2$; and gapped tertiary Dirac cones at flux $\Phi/\Phi_0=p/q$ in a region near $\mu=-\hbar vb/2$ (in the valence band only, electron-hole symmetry is strongly broken by the superlattice potential). The phases that meet at these rational flux values have Hall conductance differing by $2q e^2/h$ --- reflecting a two-fold valley degeneracy and a $q$-fold degeneracy of the magnetic minibands. (We are not counting spin.) 

Panels \ref{fig:sigmaxy}\textit{b--d} show how the connectivity switches from Fig.\ \ref{fig_transition} appear in the numerical simulation when we slightly misalign the hBN lattice relative to the graphene lattice. Each switch in the connected component of the phase diagram is associated with the closing and reopening of the Dirac cones in the magnetic Brillouin zone. (The gap closing at $\Phi=\Phi_0$ is the one shown in Fig.\ \ref{fig_transition}.) 

We will now demonstrate that transport properties near these connectivity switches have the characteristics of massless Dirac fermions \cite{Bee08}. The effects we consider are the scale-invariant (pseudodiffusive) two-terminal conductivity and sub-Poissonian shot noise at the Dirac point \cite{Kat06a,Two06}, and Klein tunneling through a potential step \cite{Kat06b,Che06}.

\begin{figure}[tb]
\centerline{\includegraphics[width=0.9\linewidth]{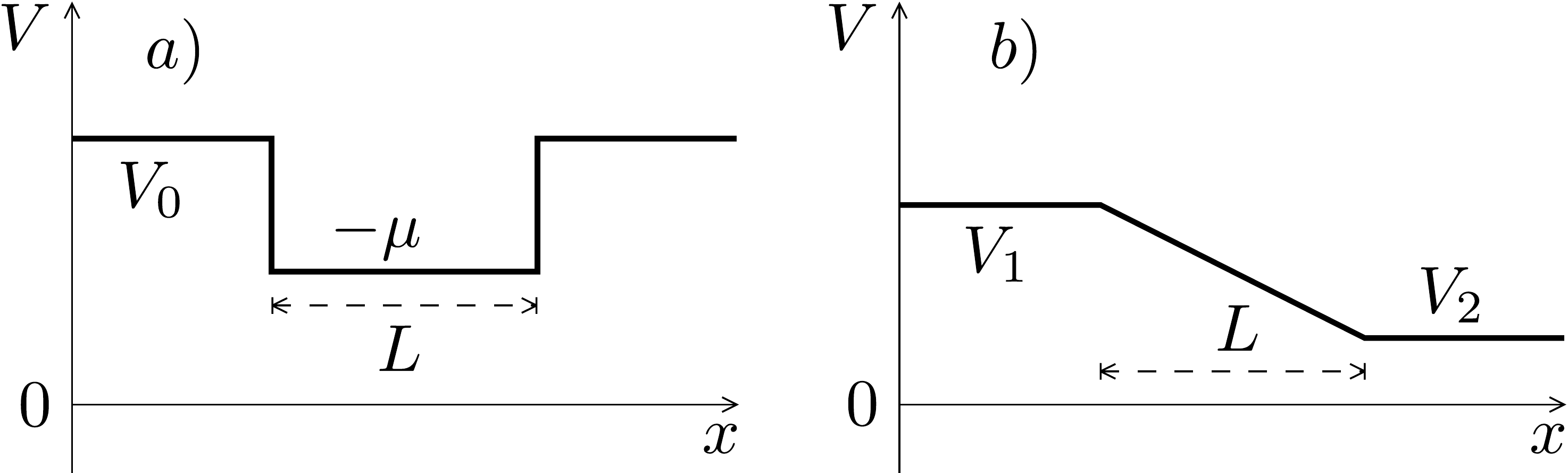}}
\caption{Electrostatic potential profile in a graphene strip, used to study the scale invariant conductivity (panel \textit{a}, $V_0/\hbar vb=1$, varying $\mu$) and Klein tunneling (panel \textit{b}, $V_1/\hbar vb=0.645$, $V_2/\hbar vb=0.613$). The Fermi level $E_{\rm F}=0$ lines up with the flux-induced Dirac point when $V\approx 0.63\,\hbar vb$.
}
\label{fig:profile}
\end{figure}

\begin{figure}[tb]
\centerline{\includegraphics[width=0.87\linewidth]{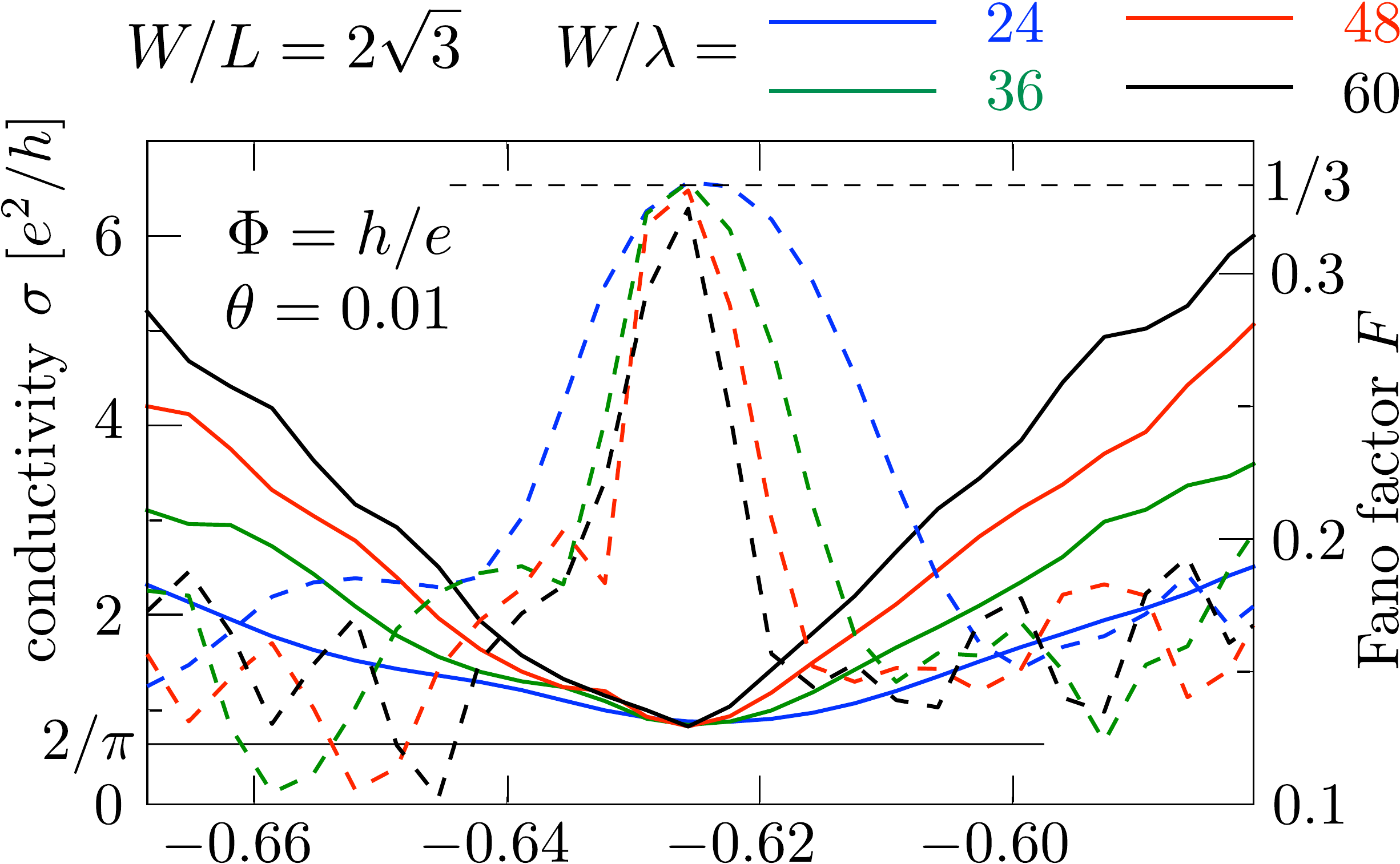}}
\caption{Conductivity (solid curves, left axis) and Fano factor (dashed curves, right axis) calculated in the two-terminal graphene strip of Fig.\ \ref{fig:profile}\textit{a}, for different system sizes at fixed aspect ratio $W/L$. The scale invariance at $\mu\approx -0.63\,\hbar vb$ signals the appearance of massless Dirac fermions at flux $\Phi=h/e$ through the superlattice unit cell. The horizontal solid and dashed lines indicate the limits \eqref{sigmaF} expected from the Dirac equation.
}
\label{fig:pseudodiv}
\end{figure}

To search for scale invariance we take an infinitely long graphene strip of width $W$, with the potential profile shown in Fig.\ \ref{fig:profile}\textit{a}. The superlattice potential is imposed over a length $L$ (where $V=-\mu$), while the leads have no superlattice ($V_{\rm lead}=\hbar vb$). The two-terminal conductivity $\sigma$ and Fano factor $F$ (ratio of noise power and current) are obtained from the transmission eigenvalues $T_{n}$,
\begin{equation}
\sigma = \frac{L}{W}\frac{e^2}{h} \sum_n T_n,\;\; F = \frac{\sum_n T_n(1-T_n)}{\sum_n T_n}.
  \label{eq:sigmaF}
\end{equation}
For $2q$ gapless Dirac cones we expect at the Dirac point the scale invariant values \cite{Kat06a,Two06} 
\begin{equation}
\sigma_{\rm D}=2qe^2/\pi h,\;\;F_{\rm D}=1/3.\label{sigmaF}
\end{equation}

We vary $W$ at fixed aspect ratio $W/L$ to search for this scale invariance. We have examined several flux values, here we show representative results for $\Phi=\Phi_0$ (so $q=1$). From Fig.\ \ref{fig:sigmaxy} we infer that the connectivity switch at this flux value happens near $\theta=0.01$ and $\mu=-0.6\,\hbar vb$. Indeed, in Fig.\ \ref{fig:pseudodiv}  both $\sigma$ and $F$ become approximately independent of sample size near these parameter values. The limiting Fano factor is close to the expected $1/3$; the limiting conductivity is a bit larger than the expected value, which we attribute to an additional contribution of order $(L/W)e^2/h$ from edge states.

\begin{figure}[tb]
\centerline{\includegraphics[width=0.9\linewidth]{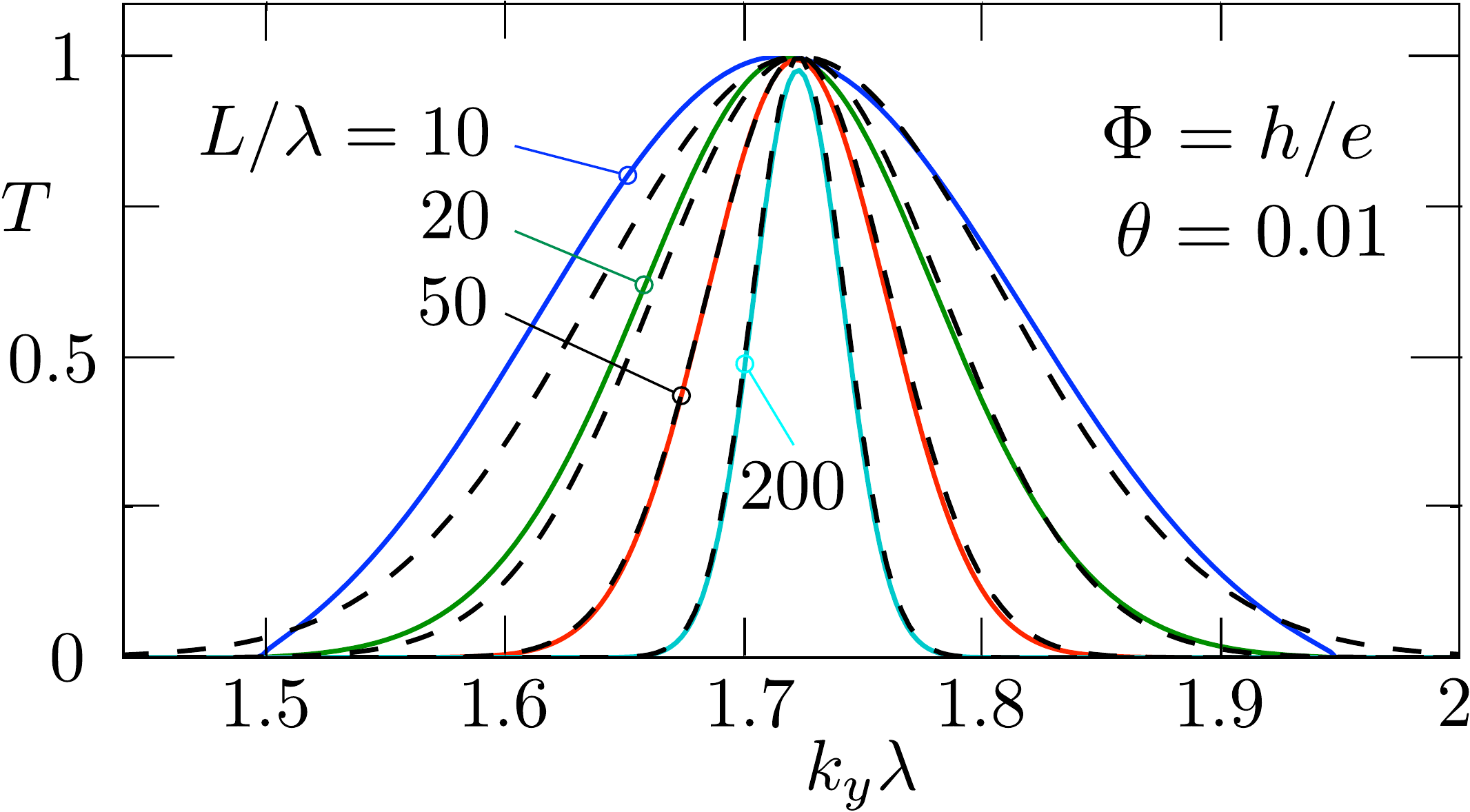}}
\caption{Transmission probability $T$ through the potential step of Fig.\ \ref{fig:profile}\textit{b}, as a function of transverse wave vector $k_y$ for different step lengths $L$. The flux-induced Dirac point is at $k_y=1.723/\lambda$. The solid curves result from the numerical simulation of the graphene superlattice at $\Phi=\Phi_0$, $\theta=0.01$, the dashed curves are the analytical prediction \eqref{eq:Tofky} for Klein tunneling of massless Dirac fermions. (There is no fit parameter in this comparison.)
}
\label{fig:klein}
\end{figure}

Klein tunneling is the transmission with unit probability at normal incidence on a potential step that crosses the Dirac point. It is a direct manifestation of the chirality of massless Dirac fermions \cite{Kat06b}. We search for this effect using the potential profile of Fig.\ \ref{fig:profile}\textit{b}, which for $\Phi=\Phi_0$ and $\theta=0.01$ is symmetrically arranged around the flux-induced Dirac point. In order to avoid spurious reflections from the leads we now apply the superlattice potential and the magnetic field to an unbounded graphene plane. We calculate the transmission probability $T(k_y)$ as a function of transverse wave vector $k_y$ in the magnetic Brillouin zone.

The dependence on the angle of incidence $\phi$ of the transmission probability of massless Dirac fermions depends exponentially on the step length $L$ \cite{Che06},
\begin{equation}
 T(\phi) = \exp(-\pi \hbar^{-1}p_{\rm F} L\sin^2\phi),
  \label{eq:Tofphi}
\end{equation}
for a symmetric junction with the same Fermi momentum $p_{\rm F}$ at both sides of the potential step. (The step should be smooth on the scale of the lattice constant, so $L\gg\lambda$ is assumed.) The transverse momentum appearing in the Dirac equation is measured from the Dirac point, $p_y=\hbar(k_y-K_y)$. (The flux $\Phi=\Phi_0$ creates two Dirac cones, both with the same value of $K_y$.) Inspection of the band structure gives  $K_y=1.723/\lambda$ and Fermi velocity $v_{\rm F}=2.04\,v$, nearly twice the native Fermi velocity $v$ of graphene. The angle of incidence then follows from $\sin\phi=p_y/p_{\rm F}$, with $p_{\rm F}=0.23\,\hbar/\lambda$, so we expect a transmission peak described by
\begin{equation}
T(k_y) = \exp(\pi \hbar L(k_y-K_y)^2/p_{\rm F}).
\label{eq:Tofky}
\end{equation}
The resulting curves are shown in Fig.\ \ref{fig:klein} (dashed curves), for different values of $L$. There is a good agreement with the numerical simulations (solid curves). 

The angle-resolved detection in these simulations is convenient to directly access the strongly peaked transmission profile \eqref{eq:Tofky}. Experimentally this signature of Klein tunneling can be observed without requiring angular resolution in a double potential step geometry \cite{You09}.

In summary, we have identified a mechanism for the production of massless Dirac fermions in the Hofstadter butterfly spectrum of a moir\'{e} superlattice. Generically, the flux-induced clones of the zero-field Dirac cones are gapped, but the gap closes at a switch in the connected component of the quantum Hall phase diagram. We have presented a model calculation for graphene on an hexagonal boron nitride surface that exhibits these connectivity switches upon variation of the crystallographic misalignment. Only a slight misalignment is needed, on the order of $1^{\circ}$, comparable to what has been realized in experiments \cite{Pon13,Dea13,Hun13,Woo14}. Numerical simulations of transport properties at unit flux through the superlattice unit cell reveal the scale invariant conductivity and Klein tunneling that are the characteristic signatures of ballistic transport of massless Dirac fermions. These should be observable in small samples, in larger samples the effects of disorder remain as an interesting problem for further research.

This research was supported by the Foundation for Fundamental Research on Matter (FOM), the Netherlands Organization for Scientific Research (NWO/OCW), an ERC Synergy Grant, and the German Academic Exchange Service (DAAD).

\appendix
\section{Derivation of the tight-binding Hamiltonian for the moir\'{e} superlattice}
\label{app:lattice-model}

Our numerical simulations are based on a tight-binding discretization of the moir\'{e} superlattice Hamiltonian \eqref{eq:model} for graphene on an hexagonal substrate. Here we provide a derivation of the tight-binding Hamiltonian, arriving at Eq.\ \eqref{eq:tbham}. This is not quite straightforward, because of the need to accomodate two lattices, of graphene and of the substrate, in a single discretization. We start with zero magnetic field ($\bm{A}=0$).

In order to achieve a commensurate discretization of the bare graphene Hamiltonian \eqref{H0def} and the moir\'e superlattice defined by reciprocal lattice vectors $\bm b_{m}(\theta)$, for arbitrary alignment angle $\theta$, we make use of the invariance of $H_0$ under a simultaneous rotation of space and pseudospin (sublattice degree of freedom).
A rotation by
\begin{align}
  -\phi = -\arctan\left(\frac{\sin\theta}{\cos\theta-(1+\delta)}\right)
\end{align}
leaves $H_0$ invariant,
\begin{equation}
v\bm p\cdot\bm\sigma+V(\bm{r})\mapsto v\tilde{\bm p}\cdot\tilde{\bm\sigma}+\tilde{V}(\tilde{\bm{r}}),\label{H0rotated}
\end{equation}
while bringing the reciprocal lattice vectors in alignment with $\bm b_{m}(\theta = 0)$. 

The first two terms of the moir\'e modulation transform into
\begin{align}
  &\hbar vbU_1 f_+[\bm r(\tilde x,\tilde y)] + i\xi\hbar vbU_2 f_-[\bm r(\tilde x,\tilde y)]\sigma_z\nonumber\\
  &\quad\quad = 
  \hbar vbU_1 \tilde f_+(\tilde{\bm r}) + i\xi\hbar vbU_2\tilde f_-(\tilde{\bm r})\tilde\sigma_z,\label{U1U2rotated}\\
&\tfrac{1}{2}\tilde f_+(\tilde{\bm r})=\cos (\bm{g}_1 \tilde{\bm r})+\cos (\bm{g}_3 \tilde{\bm r})+\cos (\bm{g}_5 \tilde{\bm r}),\\
&\tfrac{1}{2}i\tilde f_-(\tilde{\bm r})=\sin (\bm{g}_1 \tilde{\bm r})+\sin (\bm{g}_3 \tilde{\bm r})+\sin (\bm{g}_5 \tilde{\bm r}).
\end{align}
The rotated reciprocal superlattice vectors
\begin{align}
  \bm g_1 = \frac{b}{2}
  \begin{pmatrix}
	 -\sqrt{3} \\ 1
  \end{pmatrix} ,\;
  \bm g_3 = b
  \begin{pmatrix}
	 0 \\ -1
  \end{pmatrix} ,\;
  \bm g_5 = \frac{b}{2}
  \begin{pmatrix}
	 \sqrt{3} \\ 1
  \end{pmatrix} ,
  \label{eq:g1g3g5}
\end{align}
depend on $\theta$ only in their length $b=(4\pi/\sqrt{3}a)\sqrt{\delta^2+\theta^2}$, but unlike $\bm b_{m}$ not in their direction.

The third term of the moir\'{e} modulation transforms into
\begin{widetext}
\begin{align}
	 i\xi\hbar vU_3\left[-\frac{f_-[\bm r(\tilde x,\tilde y)]}{\partial \tilde y}\frac{\partial y}{\partial \tilde y}\left(\tilde \sigma_x\cos\phi-\tilde\sigma_y\sin\phi\right)+\frac{f_-[\bm r(\tilde x,\tilde y)]}{\partial \tilde x}\frac{\partial x}{\partial \tilde x}\left(\tilde\sigma_y\cos\phi+\tilde\sigma_x\sin\phi\right)\right] 
	= \xi{\mathcal A_x}(\tilde{\bm r})\tilde\sigma_x +
	\xi{\mathcal A_y}(\tilde{\bm r})\tilde\sigma_y.\label{U3rotated}
\end{align}
We have introduced the fictitious vector potential 
\begin{align}
  {\bm{\mathcal A}}(\tilde{\bm r}) = 
  \begin{pmatrix}
	 {\mathcal A_x}(\tilde{\bm r}) \\ {\mathcal A_y} (\tilde{\bm r})
  \end{pmatrix} =
  -\hbar vbU_3
  \begin{pmatrix}
	 \cos(\bm g_1\tilde{\bm r}) + \cos(\bm g_5\tilde{\bm r}) -2\cos(\bm g_3\tilde{\bm r}) \\
	 \sqrt{3}[\cos(\bm g_1\tilde{\bm r}) - \cos(\bm g_5\tilde{\bm r})]
  \end{pmatrix} \,.
  \label{eq:Atilde}
\end{align}
\end{widetext}

The full Hamiltonian in the rotated basis reads
\begin{align}
  \tilde H ={}& v\tilde{\bm p}\cdot\tilde{\bm \sigma} +\tilde{V}(\tilde{\bm{r}})+ \xi\hbar vbU_1\tilde f_+(\tilde{\bm r}) + i\xi\hbar vbU_2\tilde f_-(\tilde{\bm r})\tilde\sigma_z \nonumber\\
  &+ \xi{\bm{\mathcal A}}(\tilde{\bm r})\cdot\tilde{\bm \sigma}\;.
  \label{eq:H_tilde}
\end{align}
In the following we will work in this rotated basis, but in favor of a simple notation we will drop the tilde~$\tilde{}\,$.

\begin{figure}[h]
  \begin{center}
	 \includegraphics[width=0.9\linewidth]{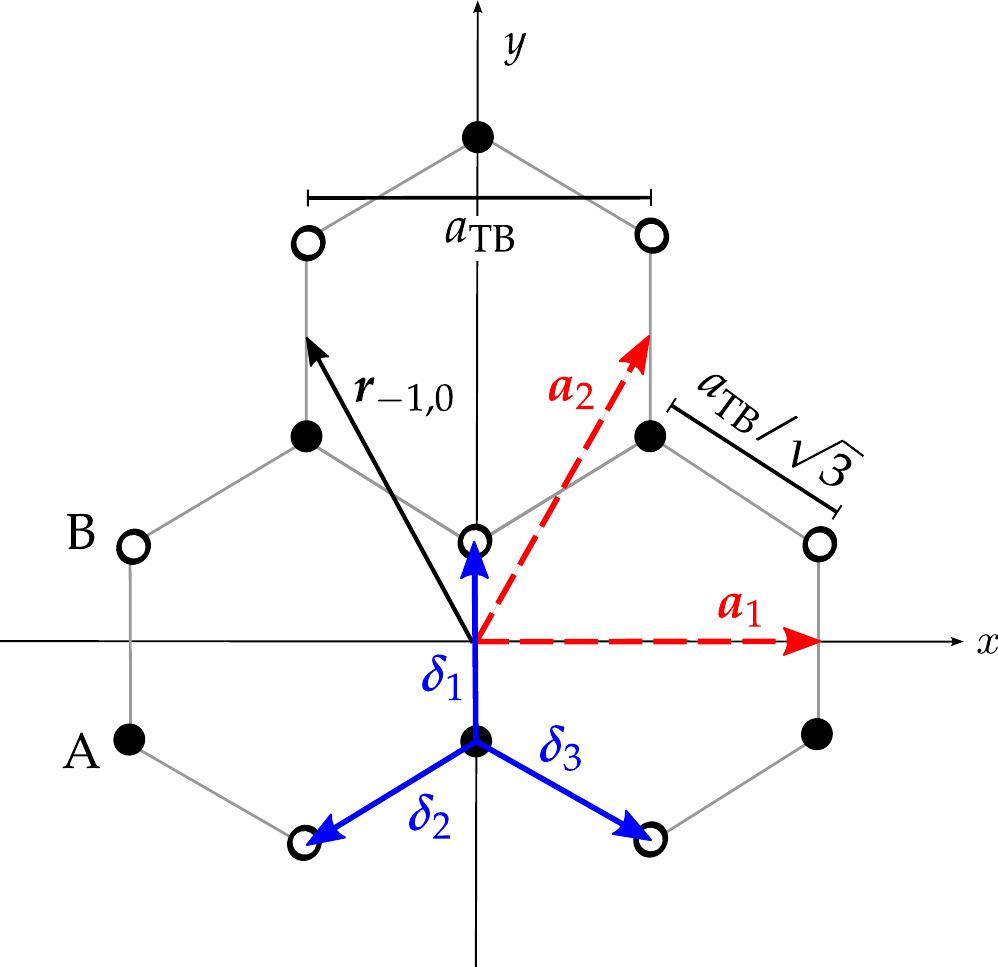}
  \end{center}
  \caption{Hexagonal lattice of the tight-binding model, with lattice vectors $\bm a_1$, $\bm a_2$ and nearest-neighbor displacement vectors $\bm\delta_1$, $\bm\delta_2$, $\bm\delta_3$. The two sublattices have sites labeled A (filled dots) and B (open dots). The vector $\bm r_{ij}=i\bm a_1+j\bm a_2$ denotes the center of unit cell $(i,j)$.}
  \label{fig:lattice}
\end{figure}

We discretize the Hamiltonian \eqref{eq:H_tilde} in the rotated basis on the hexagonal lattice, defined by the lattice vectors
\begin{align}
  \bm a_1 = a_{\rm TB}
  \begin{pmatrix}
	 1 \\ 0
  \end{pmatrix} \;,\;
  \bm a_2 = \tfrac{1}{2}a_{\rm TB}
  \begin{pmatrix}
	 1 \\ \sqrt{3}
  \end{pmatrix} \;,
  \label{eq:a1a2}
\end{align}
and the three nearest neighbor displacement vectors
\begin{align}
  &\bm\delta_1 = a_{\rm TB}
  \begin{pmatrix}
	 0 \\ 1/\sqrt{3}
  \end{pmatrix}\;,\;
  \bm\delta_2 = \tfrac{1}{2}a_{\rm TB}
  \begin{pmatrix}
	 -1 \\ -1/\sqrt{3}
  \end{pmatrix},\nonumber\\
&  \bm\delta_3 = \tfrac{1}{2}a_{\rm TB}
  \begin{pmatrix}
	 1 \\ -1/\sqrt{3}
  \end{pmatrix}\;.
  \label{eq:delta123}
\end{align}
The vector $\bm r_{ij}=i\bm a_1+j\bm a_2$, with $i,j$ integer, denotes the center of unit cell $(i,j)$.
As shown in Fig.~\ref{fig:lattice} we put the sites belonging to the A(B)-sublattice at $\bm r_{ij}-(+)\bm\delta_1/2$ to have inversion symmetry about the origin. 

To ensure that the discretization (lattice constant $a_{\rm TB}$) is commensurate with the moir\'{e} superlattice (lattice constant $\lambda$), we take an integer ratio $\lambda/a_{\rm TB}=\Lambda$, so
\begin{align}
  a_{\rm TB} = \frac{\lambda}{\Lambda} = \frac{a}{\Lambda\sqrt{\delta^2+\theta^2}}
  \label{eq:aL} .
\end{align}
The accuracy of the discretization is improved by increasing $\Lambda$. (In the simulations we take $\Lambda=20$.) 

The bare graphene Hamiltonian \eqref{H0rotated} is produced by nearest-neighbor hopping on the hexagonal lattice,
\begin{equation}
  \tilde{H}_0=-\sum_{i,j}\sum_{\alpha=1}^3 t \left[a^\dag (\bm r_{ij}^{\rm A}) b (\bm r_{ij}^{\rm A}+\bm{\delta}_\alpha) + {\rm H.c.}\right]+\sum_{i,j}\tilde{V}(\bm{r}_{ij}).\label{H0rotatedTB}
\end{equation}
Here $\bm{r}_{ij}^{\rm A}$ denotes the positions of sites on sublattice A, $a^{\dagger}$ and $b^{\dagger}$ are creation operators on the A and B sites, and $t$ is the hopping amplitude,
\begin{align}
  t = \frac{2v}{\sqrt{3}a_{\rm TB}} = \frac{2v}{\sqrt{3}a}\Lambda\sqrt{\delta^2+\theta^2} \,.
  \label{eq:t}
\end{align}

The superlattice term $U_1$ in Eq.\ \eqref{U1U2rotated} corresponds to a periodic spatial modulation of the on-site energy, the same for A and B sites, while the term $U_2$ has an additional staggering --- acting on A and B sites with opposite sign. To maintain the spatial inversion symmetry of the continuum model we evaluate both terms at the center of each unit cell. The resulting terms are given in Eqs.~\eqref{eq:even} and \eqref{eq:odd}.

The superlattice term $U_3$ with the fictitious vector potential in Eq.\ \eqref{U3rotated} represents a periodic spatial modulation of the nearest-neighbor hopping amplitudes in the tight-binding Hamiltonian \eqref{H0rotatedTB}. The replacement $t\mapsto t+\delta t_{\alpha}(\bm{r}_{ij})$ produces in the continuum limit the vector potential \cite{Cas09}
\begin{align}
  \mathcal A(\bm r) &= \sum_{\alpha=1}^3 \delta t_{\alpha}(\bm r) e^{-i\bm K \bm \delta_\alpha}= \mathcal A_x(\bm r) + i\mathcal A_y(\bm r) \label{eq:AK} \,.
\end{align}
The vectors $\bm K = (4\pi/3a_{\rm TB})\hat{x}$ and $-\bm{K}$ locate the two Dirac cones (valleys) in the hexagonal Brillouin zone.
We seek to discretize a given fictitious vector potential on the lattice, in other words we need to invert \eqref{eq:AK}.
The complex field $\mathcal A$ is constructed from three real hoppings, so we have some freedom in choosing the $\delta t_\alpha$.
We take 
\begin{align}
  &\delta t_1 = 2\mathcal A_x/3 \;,\; \delta t_2 = \mathcal A_y/\sqrt{3} -\mathcal A_x/3,\nonumber\\
  & \delta t_3 = -\mathcal A_y/\sqrt{3} -\mathcal A_x/3 \,.
  \label{eq:choice1}
\end{align}
To avoid a spurious breaking of inversion symmetry we evaluate  $\mathcal A$ in the middle of each bond, rather than on the lattice site. 

Collecting results, we arrive at the tight-binding Hamiltonian
\begin{widetext}
\begin{align}
  H ={}& \sum_{i,j} \left[ (\epsilon^{i,j}_++\epsilon^{i,j}_-+\tilde{V}(\bm{r}_{i,j}))a^\dagger_{i,j}a_{i,j} + (\epsilon^{i,j}_+-\epsilon^{i,j}_-+\tilde{V}(\bm{r}_{i,j}))b^\dagger_{i,j}b_{i,j}\right] \nonumber\\
  &- \sum_{i,j}\left[
  t_1^{i,j} a^\dagger_{i,j}b_{i,j} + t_2^{i,j}a^\dagger_{i,j}b_{i,j-1} + t_3^{i,j} a^\dagger_{i,j}b_{i+1,j-1} +{\rm H.c.}\right].
  \label{eq:tbham}
\end{align}
The energies
\begin{align}
  \epsilon_+^{i,j} &= \frac{E_0}{\hbar vb}\frac{ \delta}{\sqrt{\delta^2+\theta^2}}\frac{2\pi}{\Lambda} \left[\cos(\bm g_1\bm r_{i,j}) +\cos(\bm g_3\bm r_{i,j}) +\cos(\bm g_5\bm r_{i,j}) \right]\;, 
  \label{eq:even}\\
  \epsilon_-^{i,j} &=\frac{E_0}{\hbar vb} \frac{-\sqrt{3} \delta}{\sqrt{\delta^2+\theta^2}}\frac{2\pi}{\Lambda} \left[\sin(\bm g_1\bm r_{i,j}) +\sin(\bm g_3\bm r_{i,j}) +\sin(\bm g_5\bm r_{i,j}) \right] ,
  \label{eq:odd}
\end{align}
\end{widetext}
correspond to the periodic on-site contributions of the moir\'e super-lattice potential which are symmetric ($\epsilon_+^{i,j}$) and antisymmetrc ($\epsilon_-^{i,j}$) with respect to a swap of the A and B sublattice. 
The hoppings
\begin{subequations}
\label{eq:t1t2t3}
\begin{align}
  t_1^{i,j} ={}& t - 2\mathcal A_x(r_{i,j})/3, \\
  t_2^{i,j} ={}& t - \mathcal A_y(r_{i,j}-\delta_1/2+\delta_2/2)/\sqrt{3} \nonumber\\
  &+ \mathcal A_x(r_{i,j}-\delta_1/2+\delta_2/2)/3, \\
  t_3^{i,j} ={}& t + \mathcal A_y(r_{i,j}-\delta_1/2+\delta_3/2)/\sqrt{3}\nonumber\\
  & + \mathcal A_x(r_{i,j}-\delta_1/2+\delta_3/2)/3,
\end{align}
\end{subequations}
include both the isotropic contribution $t$ of native graphene and the periodic modulation from the moir\'e superlattice, produced by the fictitious vector potential
\begin{align}
  &\bm{\mathcal A}(\bm r) = 
  \begin{pmatrix}
  	\mathcal A_x(\bm r) \\ \mathcal A_y (\bm r)
  \end{pmatrix}  =\frac{E_0}{\hbar vb}
  \frac{-\delta^2}{\delta^2+\theta^2}\frac{2\pi}{\Lambda}\nonumber\\
  &\quad\mbox{}\times
  \begin{pmatrix}
	 \cos(\bm g_1\bm r) + \cos(\bm g_5\bm r) -2\cos(\bm g_3\bm r) \\
	 \sqrt{3}[\cos(\bm g_1\bm r) - \cos(\bm g_5\bm r)]
  \end{pmatrix} \,.
  \label{eq:A}
\end{align}

Finally, the orbital effect of the magnetic field $\bm B=B\hat{z}$ is included by adding a Peierls phase $2\pi (\Phi/\Phi_{0})\Lambda^{-2}\bm r_{i,j}\cdot\hat x$ to the hopping amplitude $t_1^{i,j}$, where $\Phi$ is the flux through the superlattice unit cell.

\begin{thebibliography}{99}
\bibitem{note1} The valley-isotropic Dirac Hamiltonian \eqref{H0def} acts on the spinor $(\Psi_{\rm A},\Psi_{\rm B})$ in valley $K$ and $(\Psi_{\rm B},-\Psi_{\rm A})$ in valley $K'$, where $\Psi_{\rm A},\Psi_{\rm B}$ are the wave amplitudes on the two triangular sublattices that form the hexagonal lattice of graphene.
\bibitem{note2} Because the spin degree of freedom is not counted, the conductance quantum is $e^2/h$ rather than $2e^2/h$.
\bibitem{note3} Details of the calculation are given in the Appendix.
\end{thebibliography}
\end{document}